# Effects of the reaction cavity on metastable optical excitation in ruthenium-sulfur dioxide complexes

Anthony E. Phillips,[1] Jacqueline M. Cole,[1,2,3] Thierry d'Almeida,[1] and Kian Sing Low[1]

[1]*Department of Physics, Cavendish Laboratory, University of Cambridge, J J Thomson Avenue, Cambridge CB3 0HE, United Kingdom*
[2]*Department of Chemistry, University of New Brunswick, P.O. Box 4400, Fredericton, New Brunswick, Canada E3B5A3*
[3]*Department of Physics, University of New Brunswick, P.O. Box 4400, Fredericton, New Brunswick, Canada E3B5A3*

We report photoexcited-state crystal structures for two new members of the $[Ru(SO_2)(NH_3)_4X]Y$ family: **1**: $X = H_2O$, $Y = (\pm)$-camphorsulfonate$_2$; **2**: $X =$ isonicotinamide, $Y =$ tosylate$_2$. The excited states are metastable at 100 K, with a photoconversion fraction of 11.1(7)% achieved in **1**, and 22.1(10)% and 26.9(10)% at the two distinct sites in **2**. We further show using solid-state density-functional-theory calculations that the excited-state geometries achieved are strongly influenced by the local crystal environment. This result is relevant to attempts to rationally design related photoexcitation systems for optical data-storage applications.

## I. INTRODUCTION

Materials with optically accessible metastable states have recently attracted attention for their potential uses in optical data storage. Their functionality arises directly from their ability to act as binary switches: if the ground state is taken to signify a "0" and the metastable state a "1," data can be written and reread using suitable wavelengths of light.[1] Photorefractive properties arising from the existence of metastable states, rather than the Pockels effect,[2] give rise to unusual and potentially useful recording kinetics.[3] However, to date relatively few materials suitable for this purpose have been identified.

One potential source of suitable materials is the field of linkage isomerism complexes. In these systems, photoexcitation of a metal-ligand charge-transfer band causes a rearrangement of the molecular geometry so that the system relaxes to a different electronic ground state determined by the new, metastable nuclear potential. The most extensively studied of these materials is sodium nitroprusside $(Na_2[Fe(CN)_5(NO)]\cdot 2H_2O)$, where the side-bound and oxygen-bound excited states of the NO ligand (in contrast to its nitrogen-bound ground state) have been identified by neutron[4] and x-ray[5] diffraction. This material has been the subject of density-functional-theory (DFT) calculations[6] and several reviews.[7] Furthermore, measurable phase gratings have been written in this material at low[8] and ambient temperatures.[9]

Similar behavior is also known in other coordination complexes. For example, the general class of compounds $[Ru(SO_2)(NH_3)_4X]Y$ also exhibits a side-bound and an end-bound metastable state (Fig. 1), studied initially in IR spectra[10] and later via x-ray diffraction and DFT calculations.[11–13]

If these or related compounds are to be useful in optical data storage, then controlled, ideally complete, conversion to their photoisomers is important for ease of reading. However, only two recently reported systems have approached 100% conversion.[14,15] Unlike spin-crossover systems, for instance, there is little evidence of cooperativity between excited centers. Various factors are known to contribute to the low conversion achieved, including limited optical penetration and the confines imposed by a relatively rigid crystal structure on the nuclear motion.[16] Comparison of gas phase with solid-state calculations has previously shown that in NO complexes the crystal surroundings have a small but important influence over the energy landscape of isomerization, in particular, influencing the barriers associated with rotation about the metal-ligand axis in the side-bound state.[17]

In this paper we focus on the triatomic ligand $SO_2$, in which the additional, free oxygen atom greatly increases the likelihood of steric interactions between the side-bound state and its crystal surroundings. The "reaction cavity" in which the ligand rotates is capable either of increasing or decreasing observed photoexcitation levels, as work on $NO_2$ complexes has demonstrated.[18] The dynamics of photoexcitation depend on the interplay between the energy costs of distorting the local and long-range structures. It is known that isomerization can cause sufficient strain on the lattice to crack a crystal.[19] We show that, on the other hand, the lattice in turn constrains the specific excited-state geometry seen in any particular case.

## II. TARGET COMPOUNDS

Members of the $[Ru(SO_2)(NH_3)_4X]Y$ series investigated in previous photocrystallographic experiments have *trans* ligands $X = H_2O$, $Cl^-$, or triflate $(CF_3CO_2^-)$ and counterions $Y = Cl^-$, benzenesulfonate $(C_6H_5SO_3^-)$, tosylate $(CH_3C_6H_4SO_3^-)$, or triflate $(CF_3CO_2^-)$.[11,12] We had the particular goal of introducing bulky substituents in the hope that these would allow the $SO_2$ ligand to rotate more freely, resulting in a higher achievable photoexcitation efficiency, and to this end varied both **X** and **Y**. Compound 1 uses the bulky ($\pm$)-camphorsulfonate (C$_{10}$H$_{15}$O$_4$S$^-$) counterion **Y**, while compound 2 uses isonicotinamide (NC$_5$H$_4$CONH$_2$) as *trans* ligand **X**, which enables the formation of rigid pairs of Ru centers joined by hydrogen bonds. Both of the compounds thus contain a newly introduced moiety **X** or **Y** and one common to previously studied complexes. Table I summarizes this and previous work. Of course, changes in the structure of individual ions, however small or systematic, do not map predictably onto changes in the crystal structure formed by packing such ions optimally. Thus the two subject complexes also represent in some sense entirely new data points from which to establish trends between photoexcitation behavior and the chemical and structural characteristics that give rise to it.

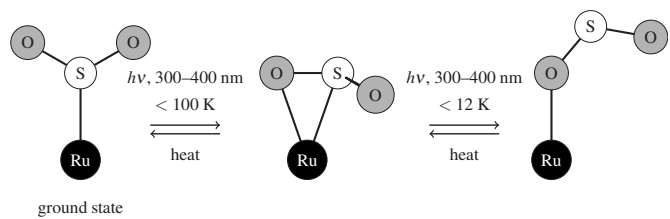

FIG. 1. Ground- and excited-state geometries known in the $[Ru(SO_2)(NH_3)_4X]Y$ family.

TABLE I. Summary of photoexcitation experiments on the [Ru(SO$_2$)(NH$_3$)$_4$**X**]**Y** family.

| X | Y | Conversion (%) | $V(SO_2)$ (Å$^{-3}$) | $T$ (K) | Reference |
|---|---|---|---|---|---|
| Cl | Cl | 10 | 41.52 | 90 | Kovalevsky *et al.*[a] |
| H$_2$O | benzenesulfonate$_2$ | 11 | 38.32 | 90 | Kovalevsky *et al.*[a] |
| H$_2$O | tosylate$_2$ | 20 | 40.32 | 90 | Kovalevsky *et al.*[b] |
| triflate | triflate | 37 | 40.33 | 90 | Kovalevsky *et al.*[b] |
| H$_2$O isonicotinamide | ($\pm$)-camphorsulfonate$_2$ | 11 | 39.25 | 100 | Present work (1) |
| | tosylate$_2$: | | | | |
| | Ru01 | 22 | 45.80 | 100 | Present work (2: Ru01) |
| | Ru51 | 27 | 45.07 | 100 | Present work (2: Ru51) |

[a]Reference 11.
[b]Reference 12.

### III. EXPERIMENTAL PROCEDURE AND RESULTS

Single-crystal x-ray diffraction data on these compounds were collected on beamline I19 at the Diamond Light Source. A full ground-state data set was collected in the dark at 100 K. The crystal was then illuminated with a focused beam of light (from a tungsten lamp for 1, a xenon lamp for 2) for two hours, during which time the crystal was rotated about its mount (i.e., the $\phi$ axis). A data set for the "light" structure was subsequently collected, using the same parameters as the "dark" collection. A Fourier difference map was used to compare the light data with the dark structural model to reveal the extent of excitation and the geometry of the excited state. The excited-state structure was subsequently refined using a nonlinear least-squares procedure.[20]

The crystallographic asymmetric unit of 1 contains a single ruthenium center, which in turn displays a single excitation geometry [Fig. 2(a)]. The excited-state atoms were refined anisotropically and their occupancy allowed to vary; this gave a photoconversion fraction of 11.1(7)%.[21]

By contrast, there are two crystallographically distinct Ru-SO$_2$ groups in structure 2, which we label Ru01 and Ru51. Of these, Ru51 displayed a single excited-state geometry **D** [Fig. 2(c)] while the excited state of Ru01 exhibited disorder about three sites **A–C** [Fig. 2(b)]. The total occupancy of the various SO$_2$ configurations was constrained (at Ru51) or restrained (at Ru01) to be 100%. The excited-state atoms were modeled isotropically, with all S atoms, bound O atoms, and free O atoms, respectively, restrained to have identical atomic displacement parameters. Finally, the two atoms comprising the side-bound ($\eta^2$) linkage were constrained to occupy the same positions regardless of which is sulfur and which is oxygen—for instance, the sulfur atom in metastable geometry **A** occupies exactly the same position as an oxygen atom in geometry **B** (Fig. 2). Although this constraint does not follow from bonding considerations in the gas phase, the electron clouds of these atoms overlap to such a considerable extent that they cannot meaningfully be resolved by these data, and attempting to do so renders the model unstable to an iterative refinement.

The three excited state geometries at Ru01 refined to a total occupancy of 11.6+7.5+3.3=22.4(10)% while the excited state geometry at Ru51 had occupancy 26.9(10)%; this is similar to excitation levels previously observed in related systems (Table I; see also Ref. 13). In general, as for NO complexes, the photoexcited state population will depend on anisotropy introduced during the irradiation process, such as the polarization and orientation of the light.[15,22] However, this effect is unlikely to be significant in this experiment due to the use of an unpolarized source and rotation about $\phi$ during irradiation and because the two Ru-SO$_2$ moeties are almost exactly antiparallel to one another (Ru-S vectors 175.0° apart, RuSO$_2$ planes 13.5° apart).

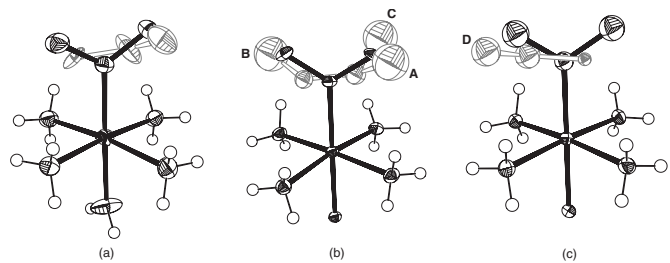

FIG. 2. Ground- and excited-state structures of (a) the Ru center in (**1**) and the two distinct centers in (**2**). Of these, (b) Ru01 exhibits disorder over three possible sites **A**, **B**, and **C** for the free O atom while (c) Ru51 displays only one orientation of the excited ligand. Atomic positions for the ground state are shown in black, for the metastable state in gray. For clarity, of the isonicotinamide ligand only the N atom bound to the Ru is shown.

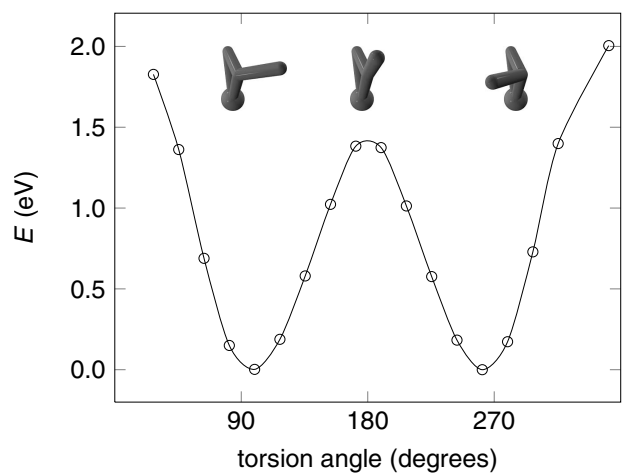

FIG. 3. Relative energies of geometry-optimized configurations of the [Ru(SO$_2$)(NH$_3$)$_4$(H$_2$O)]$^{2+}$ ion as a function of constrained Ru-O-S-O torsion angle. Torsion angles around zero are not physically sensible since they force the free O atom too close to the ammine ligands.

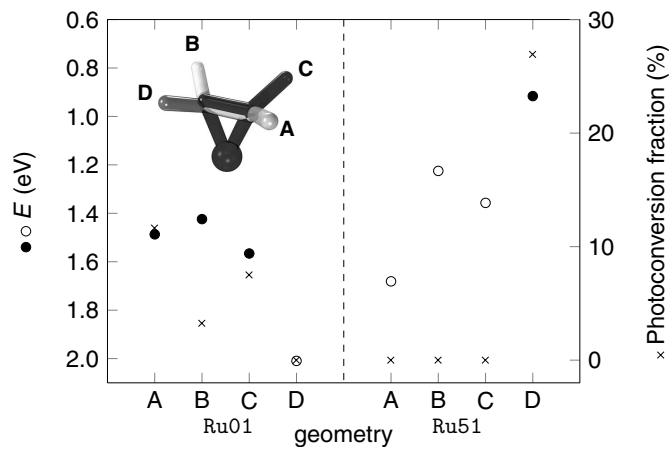

FIG. 4. DFT energies $E$ relative to the ground state (filled circles if geometries are experimentally observed, open circles if not) and experimental photoexcitation efficiencies (crosses) for the four possible excited-state geometries at each of the two Ru sites in compound 2. Note that the energy scale increases downwards so that on both vertical scales points closer to the top of the figure represent more favorable geometries.

The geometry of the excited state is also comparable with the geometry of related complexes: in particular, the S-O bond bound to the Ru is longer than the free S-O bond, as expected for bonds weakened by coordination [e.g., at Ru51 S-O$_{bound}$=1.445 Å, S-O$_{free}$=1.410 Å, mean S-O value in S-bound SO$_2$ 1.421 Å) (Ref. 23)] Atomic displacement parameters (ADPs) appear reasonable, perhaps with the exception of the relatively small ADP of the Ru-bound O atom in the metastable state at Ru51 [Fig. 2(c)]: we attribute this to the natural difficulty in partitioning electron density from the close ground-state S atom neighbor. We do not believe that this is evidence for excitation to geometries **A** or **C** at this site, as there is no electron density peak attributable to the free O atom in these geometries. The highest peaks (lowest troughs) in the Fourier difference map calculated from the final excited-state models were 1.970(−1.181) Å$^{-3}$ for compound 1 and 1.879(−2.042) Å$^{-3}$ for compound 2. These values are comparable to literature values for related systems.[12] The only residual peaks observed thought to have any physical significance were too small to be sustained in a refined model but were consistent with a small amount of rotational disorder of the SO$_3$ groups on the counterions.

## IV. DFT CALCULATIONS

In these systems the barrier to rotation about the Ru-(SO) bond (that is, to change of the N-Ru-S-O torsion angle) is known to be low so that crystal packing effects are important in determining the preferred orientation of the bound S-O linkage in the $\eta^2$ excited state.[12,17] As a result, we expect to find two local energy minima, 180° apart, for the bound S and O atoms. Associated with each of these are a further two minima for the free O atom, as expected by the pseudosymmetry of the geometry. This gives a total of four local minima for the free O atom, related approximately by reflection in the Ru-S-O plane and 180° rotation about the Ru-(SO) bond. Thus in compound 2 at Ru01 three of these possible positions are visible while only one is visible at the Ru51 site and in compound 1.

This analysis was confirmed by DFT calculations using CASTEP 5.0.1, academic release.[24] Starting from the optimized excited-state configuration of the [Ru(SO$_2$)(NH$_3$)$_4$(H$_2$O)]$^{2+}$ ion (as in 1), preliminary gas-phase geometry optimizations were performed with the Ru-O-S-O torsion angle (i.e., the position of the free oxygen atom) fixed at evenly spread values over the range of physically accessible positions. The Ru-O-S angle was held constant throughout to ensure that the SO$_2$ ligand remained in its $\eta^2$ configuration. The results confirmed that, for a given orientation of the bound O and S atoms, in the absence of crystal packing effects there are two local minima with essentially equal energies at around ±99°. (Fig. 3).

We note that if these calculations are performed without fixing the Ru-O-S angle, at several torsion angles the system becomes unstable with respect to the O-bound isomer, the highest energy metastable geometry. This suggests that suitable physical restraints on the attainable values of the torsion angle—arising, for instance, from steric repulsion—may encourage or inhibit formation of this isomer in different crystal structures.

Subsequent DFT calculations were performed in the solid state in order to elucidate the reasons for the particular orientations of the excited state observed.[24] The unit-cell parameters and positions of most nonhydrogen atoms were held constant at their crystallographically determined values (from the light data) while the positions of the SO$_2$ ligands and the hydrogen atoms were optimized from the ideal gas-phase geometry. This model is clearly idealized in that it assumes 100% excitation in a single geometry at the relevant site with zero excitation at the other. However, the results show good agreement with experiment (Fig. 4): in general, the geometries with the greatest observed occupation percentage are those where the position of the free oxygen atom comes at the lowest energy cost. (The main exception is geometry **B** at site Ru01, at which comparatively little excitation was observed, despite its energy being very close to that of geometries **A** and **C**.) These results suggest that the time scale of excitation is long enough that the $SO_2$ ligand can equilibrate to the lowest energy geometry available to it. Indeed, given the broadband irradiation the metastable state will certainly have been itself excited, and in some fraction photochemically returned to the ground state, during the course of irradiation.

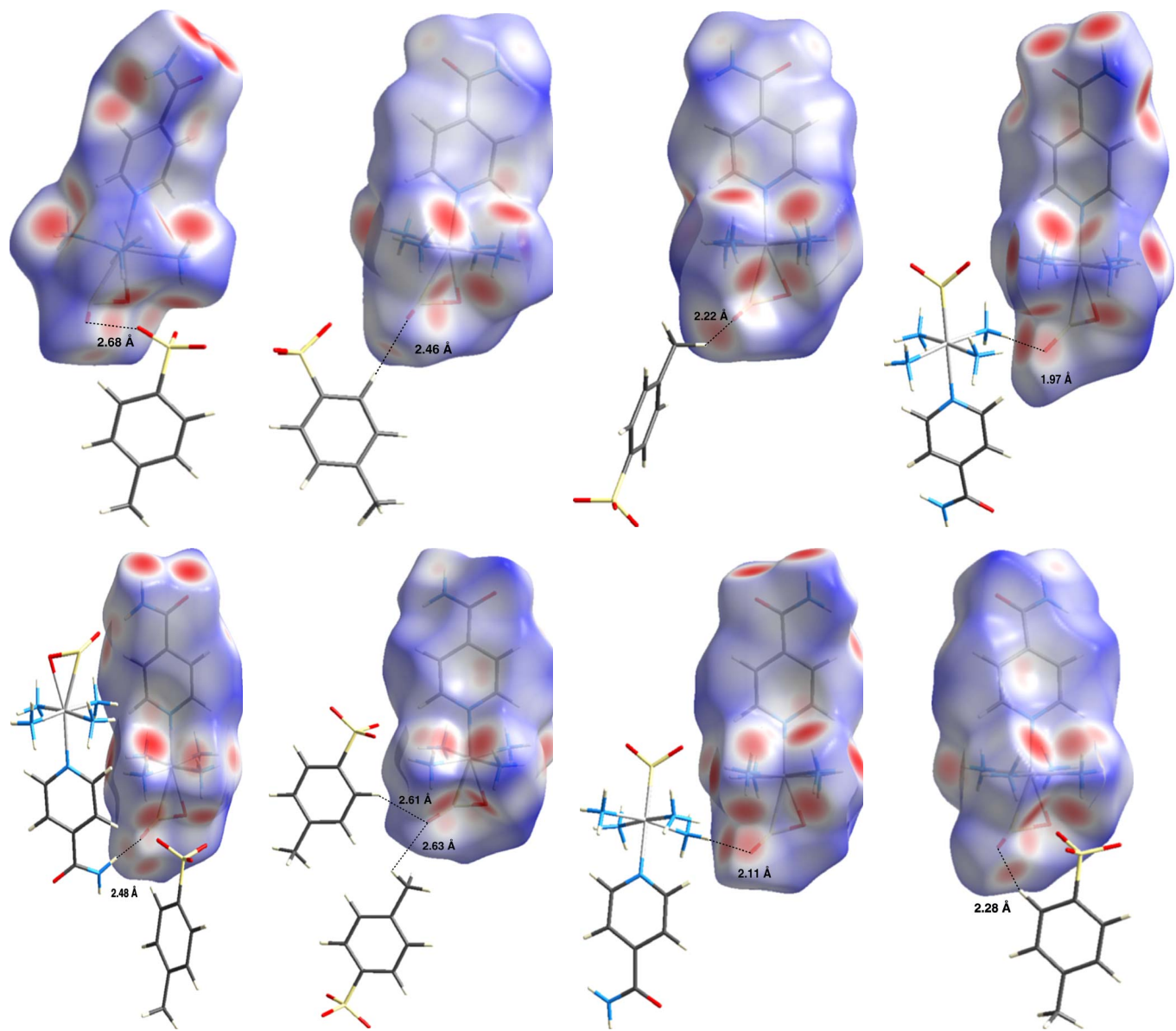

FIG. 5. (Color online) Hirshfeld surfaces for the $[Ru(SO_2)(NH_3)_4(isonicotinamide)]^{2+}$ ion in each of the four metastable geometries possible at Ru01 (above) and Ru51 (below) in compound 2, as determined by DFT calculation. Blue regions correspond to positive values of $d_{norm}$ (see definition in text), white to zero, and red to negative values. The nearest molecules to the $SO_2$ group are also shown and the closest distance between them and the free O atom indicated. The excitation percentages achieved experimentally are given for ease of comparison.

## V. GEOMETRIC CALCULATIONS

### A. Hirshfeld partitioning

The steric interactions which give rise to these energetic differences can be visualized using the Hirshfeld surfaces of the possible excited-state geometries (Fig. 5). These isosurfaces are defined as enclosing the region where the contribution due to the complex ion dominates (i.e., is at least half of) the total crystalline electron density.[25] The program CRYSTALEXPLORER 2.1 was used to plot the normalized contact distances $d_{norm}$ on these surfaces, where

$$d_{\text{norm}} = \frac{d_i - r_i^{\text{vdW}}}{r_i^{\text{vdW}}} + \frac{d_e - r_e^{\text{vdW}}}{r_e^{\text{vdW}}}.$$

Here $d_i$ is the distance to the nearest nucleus *inside* the surface and $r_i^{\text{vdW}}$ the van der Waals radius of that atom. Similarly, $d_e$ and $r_e^{\text{vdW}}$ refer to the nearest *external* nucleus. Thus regions where $d_{\text{norm}} < 0$, shown in red in Fig. 5, indicate abnormally close contact.[26]

At Ru01 (Fig. 5, top row), in the three observed geometries **A**–**C** the closest approach to the free O atom is from a hydrocarbon group with contact distance only slightly less than the sum of the van der Waals radii. (Closest approach: geometry **C**; $d = 2.22$ Å; cf. van der Waals radii[27] H = 1.20 Å, O = 1.52 Å.) In the geometry **D** which is not observed, however, the closest approach to the free O atom is an $NH_3$ group, with normalized contact distance similar to that in the hydrogen bonding between pairs of isonicotinamide ligands, visible at the top of the same diagram. ($d_{O\cdots H} = 1.97$ Å; $d_{O(\cdots H-)N} = 2.72$ Å. Amide hydrogen-bonded pair: $d_{O\cdots H} = 1.82$ Å; $d_{O(\cdots H-)N} = 2.87$ Å.) It thus appears that hydrogen bonding with the adjacent nitrogen atom distorts this geometry sufficiently to render it energetically less stable than the remaining three (Fig. 4).

At Ru51, the situation is slightly less obvious. Again, the closest contact for the observed geometry **D** is with a CH group ($d = 2.28$ Å); and again, for two of the nonobserved geometries **A** and **C**, the closest contact to the free O atom is with a NH group so that hydrogen bonding can influence the geometry. (**A**: $d_{O\cdots H} = 2.48$ Å; $d_{O(\cdots H-)N} = 3.13$ Å; **C**: $d_{O\cdots H} = 2.11$ Å; and $d_{O(\cdots H-)N} = 2.82$ Å.) However, the reason why geometry **B** is higher in energy is not clear. Indeed, of the three geometries *not* observed at this site it is the lowest in energy. Nonetheless, comparison of Fig. 5 with the DFT results in Fig. 4 show that the energy differences obtained can indeed largely be attributed to the interactions of the free O atom in the metastable state with its crystal surroundings.

### B. Voronoi-Dirichlet partitioning

Our attempts to increase the net volume of the reaction cavity—the space available for the $SO_2$ ligand to rotate in—met with only modest success. We used the program TOPOS 4.0 to partition space into Voronoi-Dirichlet polyhedra so that each point is assigned to the atom closest to it.[28] This gives a volume for the $SO_2$ group of 39.25 Å$^3$ in compound 1, which is very similar to that in previously described members of this family (Table I). Compound 2, on the other hand, does have a significantly increased volume ($\sim 45$ Å$^3$) at both of its $SO_2$ sites. However, examination of the crystal structure shows that this appears to be due to the rigid rodlike structural elements formed when hydrogen bonds between pairs of amide groups join two ruthenium complexes. It is of course notoriously difficult to predict the effects of changes in a molecule's structure on its crystal packing. Nonetheless, these results suggest a better tactic might be to embed the photoactive centers within a rigid framework, rather than attempting to surround them by poorly packing fragments. It is also worth noting that this increased volume does not lead necessarily to a greater excitation fraction; as shown in the previous section, specific intermolecular interactions are more important than net volume in influencing the excitation.

### VI. CONCLUSIONS

Photoinduced isomerism has been experimentally observed in two new complexes in the ruthenium-sulfur dioxide family: the aqua camphorsulfonate 1 and the isonicotinamide tosylate 2. While the general characteristics of these excited states are similar to previously reported members of this family, the geometric detail varies substantially as a function of the local environment, even within the one compound. In particular, the two Ru sites in the isonicotinamide tosylate structure provide an "internal standard" for one another: as excitation at these sites necessarily occurs under identical experimental conditions and to chemically identical species in near-identical orientations, geometric differences can be confidently attributed to local crystal-packing effects. We have shown using solid-state DFT calculations that, although the crystal lattice is not sufficiently rigid to prevent excitation from occurring at all, it has a strong influence on the relative occupancies observed for the four possible geometries, degenerate in the gas phase, of the excited state. Our results afford further understanding of factors impacting both the photoconversion fraction and the geometric manifestation of optical excitation. These represent an important step toward the ultimate goal of being able to tailor linkage isomerism compounds to meet the technical needs of the data storage industry.

### ACKNOWLEDGMENTS

This work was carried out with the support of the Diamond Light Source, where we gratefully acknowledge the assistance of David R. Allan, Harriott Nowell, and Sarah Barnett. Calculations were performed on the Darwin supercomputer at the Cambridge High Performance Computing Facility. A.E.P. thanks Trinity College, Cambridge for an External Research Studentship. J.M.C. is indebted to the Royal Society, the Leverhulme Trust for a Research Project Grant (for Td'A), and the University of New Brunswick for the UNB Vice-Chancellor's Research Chair. K.S.L. thanks the ESPRC (Grant No. EP/P504120/1).